# ParsRec: Meta-Learning Recommendations for Bibliographic Reference Parsing

Dominika Tkaczyk
ADAPT Centre, School of Computer Science and Statistics
Trinity College Dublin, Ireland
D.Tkaczyk@gmail.com

Paraic Sheridan
ADAPT Centre, School of Computer Science and Statistics
Trinity College Dublin, Ireland
Paraic.Sheridan@adaptcentre.ie

Joeran Beel
ADAPT Centre, School of Computer Science and Statistics
Trinity College Dublin, Ireland
Joeran.Beel@adaptcentre.ie

## ABSTRACT

Bibliographic reference parsers extract metadata (e.g. *author names*, *title*, *year)* from bibliographic reference strings. No reference parser consistently gives the best results in every scenario. For instance, one tool may be best in extracting titles, and another tool in extracting author names. In this paper, we address the problem of reference parsing from a recommender-systems perspective. We propose ParsRec, a meta-learning approach that recommends the potentially best parser(s) for a given reference string. We evaluate ParsRec on 105k references from chemistry. We propose two approaches to meta-learning recommendations. The first approach learns the best parser for an entire reference string. The second approach learns the best parser for each field of a reference string. The second approach achieved a 2.6% increase in F1 (0.909 vs. 0.886, $p < 0.001$) over the best single parser (GROBID), reducing the false positive rate by 20.2% (0.075 vs. 0.094), and the false negative rate by 18.9% (0.107 vs. 0.132).

## 1 INTRODUCTION

Bibliographic reference parsing is a well-known task in scientific information extraction. In reference parsing, the input is a single reference string, formatted in a specific bibliography style (Figure 1). The output is a machine-readable representation of the input string, typically called a parsed reference. A parsed reference is a collection of metadata fields, each of which is composed of a metadata type (e.g. "year" or "conference") and value (e.g. "2018" or "RecSys"). Reference parsing is important for academic search engines and recommender systems.

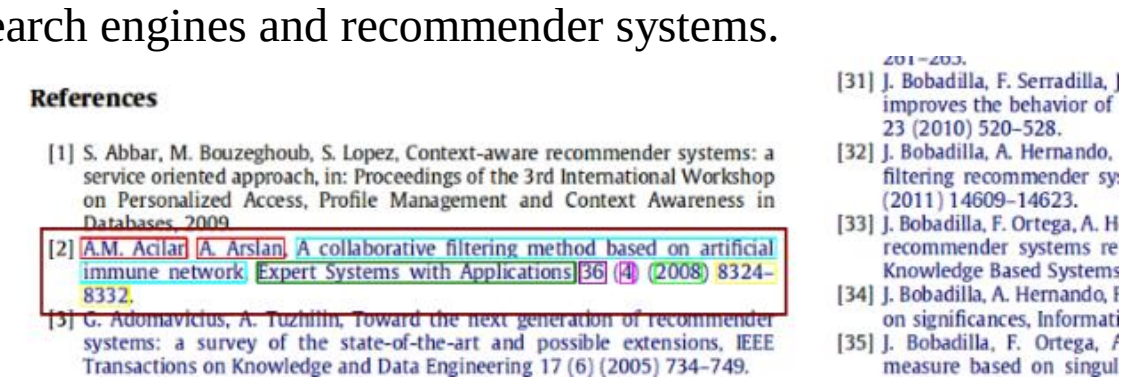
References

[1] S. Abbar, M. Bouzeghoub, S. Lopez, Context-aware recommender systems: a service oriented approach, in: Proceedings of the 3rd International Workshop on Personalized Access, Profile Management and Context Awareness in Databases, 2009.

[2] A.M. Acilar, A. Arslan, A collaborative filtering method based on artificial immune network, Expert Systems with Applications 36 (4) (2008) 8324–8332.

[3] G. Adomavicius, A. Tuzhilin, Toward the next generation of recommender systems: a survey of the state-of-the-art and possible extensions, IEEE Transactions on Knowledge and Data Engineering 17 (6) (2005) 734–749.

[31] J. Bobadilla, F. Serradilla, J improves the behavior of 23 (2010) 520–528.

[32] J. Bobadilla, A. Hernando, filtering recommender sy (2011) 14609–14623.

[33] J. Bobadilla, F. Ortega, A. H recommender systems re Knowledge Based Systems

[34] J. Bobadilla, A. Hernando, F on significances, Informati

[35] J. Bobadilla, F. Ortega, A measure based on singul

**Figure 1: An example bibliographic reference string on the input of reference parsing.**

There exist many reference parser tools, and their quality varies greatly, depending on the metadata field and other factors. For example, in our previous study [1], *ParsCit* was best in extracting authors but only third best over all fields, and *Science Parse* was best in extracting the year but only fourth best over all fields. Consequently, if we were able to choose the best parser for a given scenario, the overall quality of the results should increase. This can be seen as a typical recommendation problem: a user (e.g. a software developer) needs the item (reference parser) that satisfies the user's needs best (high-quality parsing results).

Meta-learning is often applied to the problem of algorithm selection [2]. Meta-learning allows the training of a model able to select the best algorithm for a given problem. As far as we know, meta-learning has not been applied to reference parsing.

We introduce ParsRec, a novel meta-learning approach for recommending bibliographic reference parsers. ParsRec takes as input a reference string, identifies the potentially best parser(s) for this string, applies the chosen parser(s), and outputs the extracted metadata fields. ParsRec is built upon 10 open-source parsers: Anystyle-Parser, Biblio, CERMINE, Citation, Citation-Parser, GROBID, ParsCit, PDFSSA4MET, Reference Tagger and Science Parse. ParsRec uses supervised machine learning to recommend the best parser(s). From a recommender-systems perspective, ParsRec can be seen as a switching hybrid ensemble [3] of reference parsers, where the switching is controlled by machine learning. The novel aspects of ParsRec are: 1) considering reference parsing as a recommendation problem, 2) using a meta learning approach for reference parsing.

## 2 PARSREC APPROACH

We propose and evaluate two meta-learning reference recommendation approaches, being inspired by [4].

This publication has emanated from research conducted with the financial support of Science Foundation Ireland (SFI) under Grant Number 13/RC/2016. The project has also received funding from the European Union's Horizon 2020 research and innovation programme under the Marie Skłodowska-Curie grant agreement No 713567.

ParsRec$_{Ref}$ recommends the potentially best parser for an entire reference string in three steps (Figure 2). First, for each parser, it uses a linear regression model to predict the performance of the parser (measured by F1) on the given reference string. Second, it ranks the parsers by predicted performance. Finally, it chooses the parser ranked most highly, and applies it to the input string.

ParsRec$_{Ref}$ uses two types of features to represent reference strings: 9 heuristics and 150 n-grams. The heuristics are: reference string length, number and fraction of commas, dots, and semicolons, and whether the string starts with a square bracket (e.g. "[2]"), or a dot enumeration pattern (e.g. "14."). N-gram features are 3- and 4-grams, where the terms are classes of words such as *number*, *capword* (capitalized word), *comma*, etc. These features capture style-characteristic sequences, such as *number-comma-number* (matching "3, 12"), *capword-comma-upperlett-dot* (matching "Spring, B."). The n-gram features are automatically chosen based on random forest's feature importance.

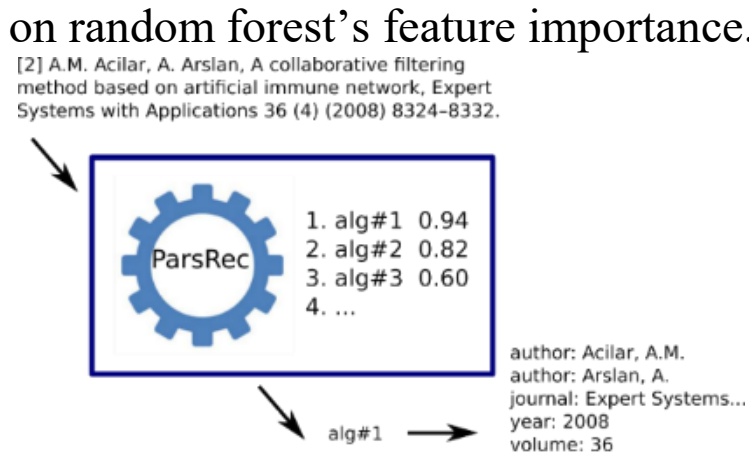

**Figure 2: The workflow of ParsRec$_{Ref}$.**

ParsRec$_{Field}$ recommends a reference parser for each metadata type in the input reference string (Figure 3). First, ParsRec$_{Field}$ iterates over all pairs (parser, metadata type), and for each pair it predicts whether the parser will correctly extract the metadata type from the input reference string. The prediction of correctness is done by a logistic regression model, trained separately for each pair (parser, metadata type). Second, for each metadata type, ParsRec$_{Field}$ ranks the parsers based on the predicted probability of correct extraction, and chooses the parser ranked most highly. All chosen parsers are applied to the input string, and the fields are chosen according to the previous choice of the parsers.

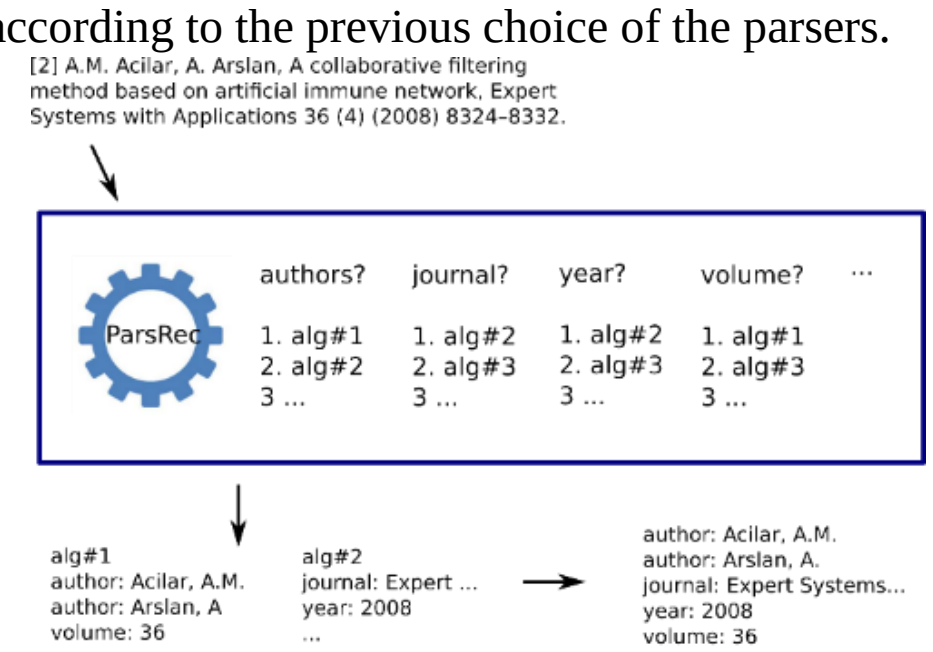

**Figure 3: The workflow of ParsRec$_{Field}$.**

## 3 EVALUATION AND RESULTS

The dataset used for the experiments comes from a business project and was manually curated. The dataset is composed of 371,656 references from chemical domains (strings and parsed versions) and 1.9 million metadata fields. The dataset contains six metadata types: *author*, *source*, *year*, *volume*, *issue*, and *page*.

The data was divided as follows: 40% for the training of individual parsers (out of scope of this paper), 30% for the training of the recommender (meta-learning), and 30% for testing.

We compare ParsRec against three baselines. The first baseline is the best single parser (GROBID). The second baseline, a hybrid baseline, uses the best parser for each metadata type separately, according to the results from [1]. The third baseline is a voting ensemble, in which the final result contains metadata fields extracted by at least three parsers. We report the results in terms of precision, recall and F1, calculated over the metadata fields.

The overall results are presented in Figure 4. Both variations of ParsRec outperform the best single parser. ParsRec$_{Ref}$ achieved a 0.6% increase in F1 (0.891 vs. 0.886), reducing the false positive rate by 3.2% (0.091 vs. 0.094), and the false negative rate by 3.8% (0.127 vs. 0.132). ParsRec$_{Field}$ achieved a 2.6% increase in F1 (0.909 vs. 0.886), reducing the false positive rate by 20.2% (0.075 vs. 0.094), and the false negative rate by 18.9% (0.107 vs. 0.132). Both increases in F1 are statistically significant (t-test, $p = 0.0027$ for ParsRec$_{Ref}$ and $p < 0.001$ for ParsRec$_{Field}$).

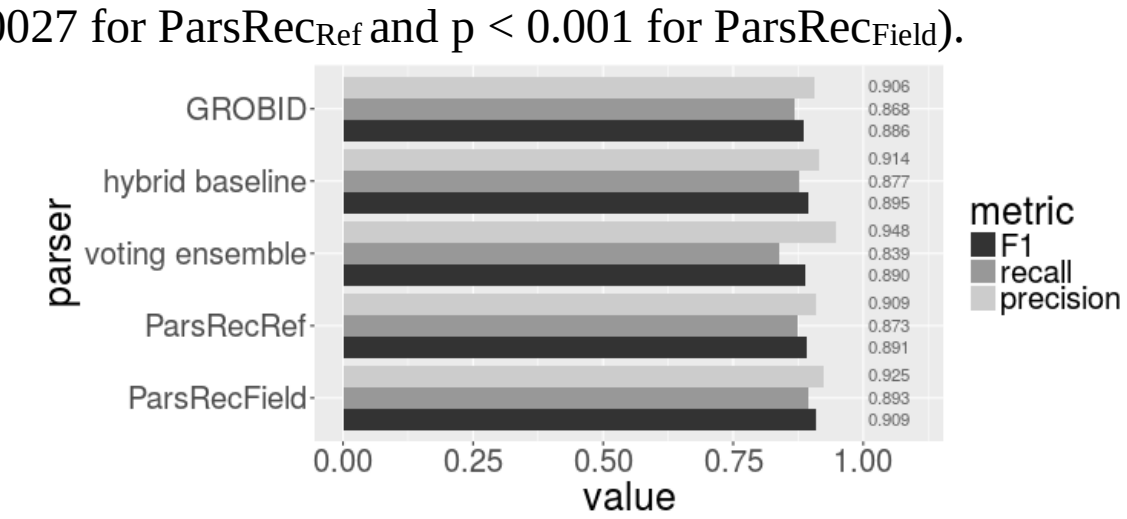

**Figure 4: The results of ParsRec and the three baselines.**

Both versions of ParsRec outperform the voting ensemble. While ParsRec$_{Ref}$ is only marginally better (F1 0.890 vs. 0.891, not significant), ParsRec$_{Field}$ achieved a 2.1% increase in F1 (0.909 vs. 0.890, $p < 0.001$). ParsRec$_{Field}$ also outperforms the hybrid baseline with a 1.6% increase in F1 (0.909 vs. 0.895, $p < 0.001$).

Our evaluation demonstrates the potential of meta learning and the application of recommendation techniques to reference parsing. Both variations of ParsRec outperform the best single parser and voting ensemble, and ParsRec$_{Field}$ outperforms all baselines. These results indicate that ParsRec makes useful recommendations. In most cases, the increases in F1 are statistically significant, though not high. We suspect the reason for this is low diversity in the data (only references from chemical papers) and among the parsers (six out of 10 parsers use Conditional Random Fields).

## REFERENCES

[1] D. Tkaczyk, A. Collins, P. Sheridan and J. Beel, "Machine Learning vs. Rules and Out-of-the-Box vs. Retrained: An Evaluation of Open-Source Bibliographic Reference and Citation Parsers," in *Joint Conference on Digital Libraries*, 2018.

[2] C. Lemke, M. Budka and B. Gabrys, "Metalearning: a survey of trends and technologies," *Artificial Intelligence Review*, vol. 44, no. 1, pp. 117-130, 2015.

[3] R. D. Burke, "Hybrid Web Recommender Systems," in *The Adaptive Web, Methods and Strategies of Web Personalization*, 2007.

[4] A. Collins, J. Beel and D. Tkaczyk, "One-at-a-time: A Meta-Learning Recommender-System for Recommendation-Algorithm Selection on Micro Level," *CoRR*, vol. abs/1805.12118.